\documentclass[12pt,titlepage]{article}
\usepackage{amsfonts}
\usepackage{amsmath}
\usepackage{amssymb}

\begin{document}

\title{Effects due to a scalar coupling on the particle-antiparticle
production in the Duffin-Kemmer-Petiau theory}
\author{T.R. Cardoso\thanks{%
cardoso@feg.unesp.br }, L.B. Castro\thanks{%
benito@feg.unesp.br }, A.S. de Castro\thanks{%
castro@pq.cnpq.br.} \\
\\
UNESP - Campus de Guaratinguet\'{a}\\
Departamento de F\'{\i}sica e Qu\'{\i}mica\\
12516-410 Guaratinguet\'{a} SP - Brazil}
\date{ }
\maketitle

\begin{abstract}
The Duffin-Kemmer-Petiau formalism with vector and scalar potentials is used
to point out a few misconceptions diffused in the literature. It is
explicitly shown that the scalar coupling makes the DKP formalism not
equivalent to the Klein-Gordon formalism or to the Proca formalism, and that
the spin-1 sector of the DKP theory looks formally like the spin-0 sector.
With proper boundary conditions, scattering of massive bosons in an
arbitrary mixed vector-scalar square step potential is explored in a simple
way and effects due to the scalar coupling on the particle-antiparticle
production and localization of bosons are analyzed in some detail. \newline
\end{abstract}

\section{Introduction}

In a recent paper (Merad 2007), scattering of massive spin-0 and spin-1
bosons under the influence of a vector smooth potential and a smooth
position-dependent mass (that is to say, a scalar smooth potential) have
been analyzed with the Duffin-Kemmer-Petiau (DKP) formalism. It has been
shown that the boundary conditions imposed on the DKP spinor for a square
step potential has to be obtained from those ones for a smooth step
potential (see also Chetouani, \textit{\ }Merad, Boudjedaa, and Lecheheb
2004), that the DKP formalism is equivalent to the Klein-Gordon and to the
Proca formalisms, and that the charge conservation law is violated under
circumstances favourable to the existence of Klein's paradox.

In the present paper, it is explicitly and precisely shown that the presence
of a scalar coupling makes the DKP formalism not equivalent to the
Klein-Gordon or to the Proca formalisms, that under the influence of scalar
and vector one-dimensional potentials the spin-1 sector of the DKP theory
looks formally like the spin-0 sector, that the proper boundary conditions
imposed on the DKP spinor for a square step potential become evident without
recurring to the limit \ of scattering in a smooth step potential and to
Heun's function. Furthermore, effects due to a scalar coupling on the
particle-antiparticle production are analyzed in some detail and it is
pointed out that the charge is always conserved if one uses an acceptable
definition of the reflection and transmission coefficients. An apparent
paradox concerning the uncertainty principle is solved by introducing the
concept of effective Compton wavelength. Comparison with the results
obtained formerly with the Klein-Gordon formalism (Cardoso and de Castro
2007) highlights the differences between the DKP and the Klein-Gordon
formalisms.

\section{The DKP equation}

The first-order DKP equation for a massive free boson is given by (Kemmer
1939)%
\begin{equation}
\left( i\beta ^{\mu }\partial _{\mu }-m\right) \psi =0  \label{dkp}
\end{equation}%
\noindent where the matrices $\beta ^{\mu }$\ satisfy the algebra%
\begin{equation}
\beta ^{\mu }\beta ^{\nu }\beta ^{\lambda }+\beta ^{\lambda }\beta ^{\nu
}\beta ^{\mu }=g^{\mu \nu }\beta ^{\lambda }+g^{\lambda \nu }\beta ^{\mu }
\label{beta}
\end{equation}%
\noindent and the metric tensor is $g^{\mu \nu }=\,$diag$\,(1,-1,-1,-1)$.
The algebra expressed by (\ref{beta}) generates a set of 126 independent
matrices whose irreducible representations are a trivial representation, a
five-dimensional representation and a ten-dimensional representation. The
second-order Klein-Gordon and Proca equations are obtained when one selects
the spin-0 and spin-1 sectors of the DKP theory. A well-known conserved
four-current is given by
\begin{equation}
J^{\mu }=\bar{\psi}\beta ^{\mu }\psi   \label{corrente}
\end{equation}%
\noindent where the adjoint spinor $\bar{\psi}=\psi ^{\dagger }\eta ^{0}$,
with $\eta ^{0}=2\beta ^{0}\beta ^{0}-1$ in such a way that $\left( \eta
^{0}\beta ^{\mu }\right) ^{\dagger }=\eta ^{0}\beta ^{\mu }$. Despite the
similarity to the Dirac equation, the DKP equation involves singular
matrices, \ the time component of $J^{\mu }$ given by (\ref{corrente}) is
not positive definite and the case of massless bosons can not be obtained by
a limiting process. Nevertheless, the matrices $\beta ^{\mu }$ plus the unit
operator generate a ring consistent with integer-spin algebra (Krajcik and
Nieto 1974) and $J^{0}$ may be interpreted as a charge density.

With the introduction of interactions, the DKP equation for a massive boson
can be written as%
\begin{equation}
\left( i\beta ^{\mu }\partial _{\mu }-m-U\right) \psi =0  \label{dkp2}
\end{equation}%
where the potential matrix $U$ with scalar and vector terms is in the form%
\begin{equation}
U=S+\beta ^{\mu }A_{\mu }  \label{pot}
\end{equation}%
with $S$ and $A_{\mu }$ denoting the scalar and four-vector potential
functions, respectively. Recently, by a proper interpretation of the DKP
spinor components, it has been shown an anomalous term already noted by
Kemmer (Kemmer 1939) disappears from the DKP formalism so that the DKP
equation and the Klein-Gordon and Proca equations are equivalent under
minimal coupling (Nowakowski 1998; Lunardi, Pimentel, Teixeira, and Valverde
2000). It is still true that $J^{\mu }$ is a conserved quantity in the
presence of interactions expressed by (\ref{pot}) and that it can be
interpreted as a charge current. \

For the case of spin 0, we use the representation for the $\beta ^{\mu }$\
matrices given by (Nedjadi and Barret 1993)%
\begin{equation}
\beta ^{0}=%
\begin{pmatrix}
\theta & \overline{0} \\
\overline{0}^{T} & \mathbf{0}%
\end{pmatrix}%
,\quad \beta ^{i}=%
\begin{pmatrix}
\widetilde{0} & \rho _{i} \\
-\rho _{i}^{T} & \mathbf{0}%
\end{pmatrix}%
,\quad i=1,2,3  \label{rep}
\end{equation}%
\noindent where%
\begin{eqnarray}
\ \theta &=&%
\begin{pmatrix}
0 & 1 \\
1 & 0%
\end{pmatrix}%
,\quad \rho _{1}=%
\begin{pmatrix}
-1 & 0 & 0 \\
0 & 0 & 0%
\end{pmatrix}
\notag \\
&&  \label{rep2} \\
\rho _{2} &=&%
\begin{pmatrix}
0 & -1 & 0 \\
0 & 0 & 0%
\end{pmatrix}%
,\quad \rho _{3}=%
\begin{pmatrix}
0 & 0 & -1 \\
0 & 0 & 0%
\end{pmatrix}
\notag
\end{eqnarray}%
\noindent $\overline{0}$, $\widetilde{0}$ and $\mathbf{0}$ are 2$\times $3, 2%
$\times $2 \ and 3$\times $3 zero matrices, respectively, while the
superscript T designates matrix transposition. The five-component spinor can
be written as $\psi ^{T}=\left( \psi _{1},...,\psi _{5}\right) $ in such a
way that the DKP equation for a boson constrained to move along the $x$-axis
decomposes into
\begin{equation*}
D_{0}\psi _{1}=-i\left( m+S\right) \psi _{2},\quad D_{1}\psi _{1}=-i\left(
m+S\right) \psi _{3}
\end{equation*}%
\begin{equation}
D_{0}\psi _{2}-D_{1}\psi _{3}=-i\left( m+S\right) \psi _{1}  \label{DKP3}
\end{equation}%
\begin{equation*}
\psi _{4}=\psi _{5}=0
\end{equation*}%
where%
\begin{equation}
D_{\mu }=\partial _{\mu }+iA_{\mu }  \label{dzao}
\end{equation}%
and $J^{\mu }$ can be written as%
\begin{equation}
J^{0}=2\,\text{Re}\left( \psi _{2}^{\ast }\psi _{1}\right) ,\quad J^{1}=-2\,%
\text{Re}\left( \psi _{3}^{\ast }\psi _{1}\right) ,\quad J^{2}=J^{3}=0
\label{corrente3}
\end{equation}%
It is worthwhile to note that $\left( D^{\mu }D_{\mu }+m^{2}\right) \psi
_{1}=0$ in the absence of the scalar potential, so that the DKP equation
reduces to the Klein-Gordon equation. The form $\partial _{1}+iA_{1}$ in Eq.
(\ref{DKP3}) suggests that the space component of the minimal vector
potential can be gauged away by defining a new spinor%
\begin{equation}
\tilde{\psi}\left( x,t\right) =\exp \left[ i\int^{x}d\zeta \,A_{1}\left(
\zeta ,t\right) \right] \psi \left( x,t\right)  \label{gauge}
\end{equation}%
Then, without loss of generality, we will consider $A_{1}=0$.

For the case of spin 1, the $\beta ^{\mu }$\ matrices are (Nedjadi and
Barret 1994)%
\begin{equation}
\beta ^{0}=%
\begin{pmatrix}
0 & \overline{0} & \overline{0} & \overline{0} \\
\overline{0}^{T} & \mathbf{0} & \mathbf{I} & \mathbf{0} \\
\overline{0}^{T} & \mathbf{I} & \mathbf{0} & \mathbf{0} \\
\overline{0}^{T} & \mathbf{0} & \mathbf{0} & \mathbf{0}%
\end{pmatrix}%
,\;\beta ^{i}=%
\begin{pmatrix}
0 & \overline{0} & e_{i} & \overline{0} \\
\overline{0}^{T} & \mathbf{0} & \mathbf{0} & -is_{i} \\
-e_{i}^{T} & \mathbf{0} & \mathbf{0} & \mathbf{0} \\
\overline{0}^{T} & -is_{i} & \mathbf{0} & \mathbf{0}%
\end{pmatrix}
\label{betaspin1}
\end{equation}%
\noindent where $s_{i}$ are the 3$\times $3 spin-1 matrices $\left(
s_{i}\right) _{jk}=-i\varepsilon _{ijk}$, $e_{i}$ are the 1$\times $3
matrices $\left( e_{i}\right) _{1j}=\delta _{ij}$ and $\overline{0}=%
\begin{pmatrix}
0 & 0 & 0%
\end{pmatrix}%
$, while\textbf{\ }$\mathbf{I}$ and $\mathbf{0}$\textbf{\ }designate the 3$%
\times $3 unit and zero matrices, respectively. \noindent In the wake of
previous works (Chetouani, \textit{\ }Merad, Boudjedaa and Lecheheb 2004;
Merad, 2007), the spinor $\psi ^{T}=\left( \psi _{1},...,\psi _{10}\right) $
can be partitioned as%
\begin{equation}
\psi _{I}^{T}=\left( \psi _{3},\psi _{4},\psi _{5}\right) ,\quad \psi
_{II}^{T}=\left( \psi _{6},\psi _{7},\psi _{2}\right) ,\quad \psi
_{III}^{T}=\left( \psi _{10},-\psi _{9},\psi _{1}\right)  \label{spinor}
\end{equation}%
so that the one-dimensional DKP equation can be expressed in the form
\begin{equation*}
D_{0}\psi _{I}=-im\psi _{II},\quad D_{1}\psi _{I}=-im\psi _{III}
\end{equation*}%
\begin{equation}
D_{0}\psi _{II}-D_{1}\psi _{III}=-i\left( m+S\right) \psi _{I}  \label{DKp3}
\end{equation}%
\begin{equation*}
\psi _{8}=0
\end{equation*}%
where $D_{\mu }$ is again given by (\ref{dzao}). In addition, expressed in
terms of (\ref{spinor}) the current can be written as
\begin{equation}
J^{0}=2\,\text{Re}\left( \psi _{II}^{\dagger }\psi _{I}\right) ,\quad
J^{1}=-2\,\text{Re}\left( \psi _{III}^{\dagger }\psi _{I}\right) ,\quad
J^{2}=J^{3}=0  \label{C1}
\end{equation}%
Comparison of (\ref{DKP3}) with (\ref{DKp3}) evidences that the spinors $%
\psi _{I}$, $\psi _{II}$ and $\psi _{III}$ behave like the spinor components
$\psi _{1}$, $\psi _{2}$ and $\psi _{3}$, respectively, from the spin-0
sector of the DKP theory. More than this, comparison of (\ref{corrente3})
with (\ref{C1}) places on view that the spin-1 sector of the DKP theory
looks formally like the spin-0 sector.

According to the observation of the last paragraph, we will restrict our
attention to the spin-0 sector of the DKP theory. If the terms in the
potential $U$ are time-independent, one can write $\psi (x,t)=\varphi
(x)\exp (-iEt)$ in such a way that the time-independent DKP equation for the
spin-0 sector splits into%
\begin{equation*}
\left( m+S\right) \frac{d}{dx}\left( \frac{1}{m+S}\,\varphi _{1}^{\prime
}\right) +K^{2}\varphi _{1}=0
\end{equation*}%
\begin{equation}
\varphi _{2}=\frac{E-A_{0}}{m+S}\,\varphi _{1}  \label{dkp4}
\end{equation}%
\begin{equation*}
\varphi _{3}=\frac{i}{m+S}\,\varphi _{1}^{\prime }
\end{equation*}%
where the prime denotes derivative with respect to $x$ and
\begin{equation}
K^{2}=\left( E-A_{0}\right) ^{2}-\left( m+S\right) ^{2}  \label{kk}
\end{equation}%
For this time-independent problem, $J^{\mu }$ has the components
\begin{equation}
J^{0}=2\,\frac{E-A_{0}}{m+S}\,|\varphi _{1}|^{2},\quad J^{1}=2\,\frac{\text{%
Im}\left( \varphi _{1}^{\prime }\varphi _{1}^{\ast }\right) }{m+S}
\label{corrente4}
\end{equation}%
Since $J^{\mu }$ is not time dependent, $\varphi $ describes a stationary
state.

Just as in the case of a pure vector coupling (Cardoso, Castro, and de
Castro 2008), there is no reason to require that the spinor and its
derivative are continuous across finite discontinuities of the potential, as
naively advocated in (Chetouani, \textit{\ }Merad, Boudjedaa and Lecheheb
2004; Merad 2007). A careful analysis reveals, though, that proper matching
conditions follow from the differential equations obeyed by the spinor
components, as they should be, avoiding in this manner to recur to the limit
process of smooth potentials. \noindent The effect of the discontinuity of
the potential can be evaluated by integrating the equations for the
components of the DKP spinor from $-\delta $ to $+\delta $, by supposing
that $x=0$ is the point of interest, and taking the limit $\delta
\rightarrow 0$. In fact, the second-order differential equation given by the
third line of (\ref{dkp4}) implies that $\varphi _{1}$ is continuous and the
first line implies that so is $\varphi _{1}^{\prime }/\left( m+S\right)
=-i\varphi _{3}$. In this case, $\varphi _{2}$ is discontinuous and so is $%
J^{0}$, but not $J^{1}$. A possible discontinuity of $J^{0}$ would not
matter if it is to be interpreted as a charge density but $J^{1}$ (involving
$\varphi _{1}^{\ast }\,\varphi _{1}^{\prime }/\left( m+S\right) $) should be
continuous in a stationary regime.

\section{The step potential}

The one-dimensional square step potential is expressed as
\begin{equation}
S=\theta \left( x\right) c_{S}V,\quad A_{0}=\theta \left( x\right) c_{A}V
\label{pot1}
\end{equation}%
\noindent where $c_{S}$ and $c_{A}$ are dimensionless and positive coupling
constants constrained by $c_{S}+c_{A}=1$, $\theta \left( x\right) $ denotes
the Heaviside step function and $V>0$ is the height of the step. For $x<0$
the DKP equation has the solution
\begin{equation}
\varphi \left( x\right) =\varphi _{+}e^{+ikx}+\varphi _{-}e^{-ikx}
\label{sol z<0}
\end{equation}%
\noindent where%
\begin{equation}
\varphi _{\pm }^{T}=\frac{a_{\pm }}{\sqrt{2}}\left( 1,\frac{E}{m},\mp \frac{k%
}{m},0,0\right)  \label{sol2}
\end{equation}%
\noindent and%
\begin{equation}
k=\sqrt{E^{2}-m^{2}}  \label{kkk}
\end{equation}%
For $\left\vert E\right\vert >m$, the solution expressed by (\ref{sol z<0})
and (\ref{sol2}) describes plane waves propagating on both directions of the
$x$-axis. The flux related to the current $J^{\mu }$, corresponding to $%
\varphi $ given by (\ref{sol z<0}), is expressed as
\begin{equation}
J^{1}=\frac{k}{m}\left( \left\vert a_{+}\right\vert ^{2}-\left\vert
a_{-}\right\vert ^{2}\right)  \label{j1}
\end{equation}%
\noindent and%
\begin{equation}
J_{\pm }^{0}=\frac{E}{m}\left\vert a_{\pm }\right\vert ^{2}  \label{j2}
\end{equation}%
\noindent If we choose incident particles on the potential barrier ($J^{0}>0
$), $\varphi _{+}\exp (+ikx)$ will describe incident particles ($J^{1}>0$),
whereas $\varphi _{-}\exp (-ikx)$ will describe reflected particles ($%
J^{1}<0 $). On the other hand, for $x>0$ the solution describes an
evanescent wave or a progressive wave running away from the potential
interface. The general solution has the form
\begin{equation}
\varphi _{\text{t}}\left( x\right) =\left( \varphi _{\text{t}}\right)
_{+}e^{+iqx}+\left( \varphi _{\text{t}}\right) _{-}e^{-iqx}  \label{sol z0}
\end{equation}%
\noindent where%
\begin{equation}
\left( \varphi _{\text{t}}\right) _{\pm }^{T}=\frac{b_{\pm }}{\sqrt{2}}%
\left( 1,\frac{E-c_{A}V}{m+c_{S}V},\frac{\mp q}{m+c_{S}V},0,0\right)
\label{sol3}
\end{equation}%
\noindent and%
\begin{equation}
q=\sqrt{\left( E-c_{A}V\right) ^{2}-\left( m+c_{S}V\right) ^{2}}  \label{q}
\end{equation}%
Due to the twofold possibility of signs for the energy of a stationary
state, the solution involving $b_{-}$ can not be ruled out a priori. As a
matter of fact, this term may describe a progressive wave with a negative
charge density and a negative flux of charge ($J^{1}<0$). In other words,
the solution $\left( \varphi _{\text{t}}\right) _{-}\exp \left( -iqx\right) $
with $q\in \mathbb{R}$ reveals a signature of Klein's paradox. One can
readily envisage that three different classes of solutions can be segregated:

\begin{itemize}
\item Class A. For $V<E-m$ one has $q\in\mathbb{R}$, and the solution
describing a plane wave propagating in the positive direction of the $x$%
-axis is possible only if $b_{-}=0$. In this case the components of the
current are given by
\end{itemize}

\begin{equation}
J^{0}=\frac{E-c_{A}V}{m+c_{S}V}\left\vert b_{+}\right\vert ^{2},\quad J^{1}=%
\frac{q}{m+c_{S}V}\left\vert b_{+}\right\vert ^{2}  \label{c11}
\end{equation}

\begin{itemize}
\item Class B. For $E-m<V<V_{c}$, where%
\begin{equation}
V_{c}=\left\{
\begin{array}{c}
\frac{E+m}{2c_{A}-1},\quad \text{for\quad }c_{A}>1/2 \\
\\
\infty ,\quad \text{for\quad }c_{A}\leq 1/2%
\end{array}%
\right.  \label{vc}
\end{equation}%
one has that $q=+i\left\vert q\right\vert $ or $q=-i\left\vert q\right\vert $%
. The solution with $q=\pm i\left\vert q\right\vert $ demands $b_{\mp }=0$
for furnishing a finite charge density as $x\rightarrow \infty $. In this
case\bigskip\ $J^{1}=0$ and%
\begin{equation}
J^{0}=\left\{
\begin{array}{c}
\frac{E-c_{A}V}{m+c_{S}V}\,e^{-2\left\vert q\right\vert x}\left\vert
b_{+}\right\vert ^{2},\quad \text{for\quad }q=+i\left\vert q\right\vert
\quad (V<E/c_{A}) \\
\\
-\frac{c_{A}V-E}{m+c_{S}V}\,e^{-2\left\vert q\right\vert x}\left\vert
b_{-}\right\vert ^{2},\quad \text{for\quad }q=-i\left\vert q\right\vert
\quad (V>E/c_{A})%
\end{array}%
\right.  \notag
\end{equation}

\item Class C. With $V>V_{c}$ it appears again the possibility of
propagation in the positive direction of the $x$-axis, now with $b_{+}=0$.
The current takes the form%
\begin{equation}
J^{0}=-\frac{c_{A}V-E}{m+c_{S}V}\left\vert b_{-}\right\vert ^{2},\quad
J^{1}=-\frac{q}{m+c_{S}V}\left\vert b_{-}\right\vert ^{2}  \label{c33}
\end{equation}
\end{itemize}

\noindent\ The demand for continuity of $\varphi _{1}$ and $\varphi
_{1}^{\prime }/\left( m+S\right) $ at $x=0$ fixes the wave amplitudes in
terms of the amplitude of the incident wave, viz. \bigskip\ \bigskip
\begin{equation}
\frac{a_{-}}{a_{+}}=\left\{
\begin{array}{c}
\frac{k-\tilde{q}}{k+\tilde{q}} \\
\\
\frac{\left( k-i|\tilde{q}|\right) ^{2}}{k^{2}+|\tilde{q}|^{2}} \\
\\
\frac{k+\tilde{q}}{k-\tilde{q}}%
\end{array}%
\begin{array}{c}
\text{\textrm{for the class A}} \\
\\
\text{\textrm{for the class B}} \\
\\
\text{\textrm{for the class C}}%
\end{array}%
\right.  \label{12}
\end{equation}%
\begin{equation}
\frac{b_{+}}{a_{+}}=\left\{
\begin{array}{c}
\frac{2k}{k+\tilde{q}} \\
\\
\frac{2k\left( k-i|\tilde{q}|\right) }{k^{2}+|\tilde{q}|^{2}} \\
\\
0%
\end{array}%
\begin{array}{c}
\text{\textrm{for the class A}} \\
\\
\text{\textrm{for the class B}} \\
\\
\text{\textrm{for the class C}}%
\end{array}%
\right.  \label{13}
\end{equation}%
\begin{equation}
\frac{b_{-}}{a_{+}}=\left\{
\begin{array}{c}
0 \\
\\
0 \\
\\
\frac{2k}{k-\tilde{q}}%
\end{array}%
\begin{array}{c}
\text{\textrm{for the class A}} \\
\\
\text{\textrm{for the class B}} \\
\\
\text{\textrm{for the class C}}%
\end{array}%
\right.  \label{14}
\end{equation}%
where $\tilde{q}=q\left( 1+c_{S}V/m\right) ^{-1}$. Now we focus attention on
the calculation of the reflection ($R$) and transmission ($T$) coefficients.
The reflection (transmission) coefficient is defined as the ratio of the
reflected (transmitted) flux to the incident flux. Since $\partial
J^{0}/\partial t=0$ for stationary states, one has that $J^{1}$ is
independent of $x$. This fact implies that \bigskip
\begin{equation}
R=\left\{
\begin{array}{c}
\left( \frac{k-\tilde{q}}{k+\tilde{q}}\right) ^{2} \\
\\
1 \\
\\
\left( \frac{k+\tilde{q}}{k-\tilde{q}}\right) ^{2}%
\end{array}%
\begin{array}{c}
\text{\textrm{for the class A}} \\
\\
\text{\textrm{for the class B}} \\
\\
\text{\textrm{for the class C}}%
\end{array}%
\right.  \label{15}
\end{equation}%
\begin{equation}
T=\left\{
\begin{array}{c}
\frac{4k\tilde{q}}{(k+\tilde{q})^{2}} \\
\\
0 \\
\\
-\frac{4k\tilde{q}}{(k-\tilde{q})^{2}}%
\end{array}%
\begin{array}{c}
\text{\textrm{for the class A}} \\
\\
\text{\textrm{for the class B}} \\
\\
\text{\textrm{for the class C}}%
\end{array}%
\right.  \label{16}
\end{equation}

\noindent It is instructive to note that (\ref{12})-(\ref{16}) look like
those ones for the mixed vector-scalar square step potential in the
Klein-Gordon formalism (Cardoso and de Castro 2007). Interestingly, $\tilde{q%
}=q\left( 1+c_{S}V/m\right) ^{-1}$ in the DKP formalism whereas $\tilde{q}=q$
in the Klein-Gordon formalism. The expression for $\tilde{q}$ departs from $%
q $ just by the factor $\left( 1+c_{S}V/m\right) ^{-1}$. It is clear that
the scalar coupling makes all the difference, even in the nonrelativistic
limit.

In the class C we meet a bizarre circumstance as long as both $J^{0}$ and $%
J^{1}$ are negative quantities. It is satisfactory to interpret the solution
$\left( \varphi _{\text{t}}\right) _{-}\exp (-iqx)$ as describing the
propagation, in the positive direction of the $x$-axis, of particles with
charges of opposite sign to the incident particles. This interpretation is
consistent if the particles moving in this region have energy $-E$ and are
under the influence of a potential $-c_{A}V$. It means that, in fact, the
progressive wave describes the propagation of antiparticles in the positive
direction of the $x$-axis. For all the classes one has \ $R+T=1$ as should
be expected for a conserved quantity. The class C presents $R>1$, the
alluded Klein's paradox, implying that more particles are reflected from the
potential barrier than those incoming. It must be so because, as seen
before, the potential stimulates the production of antiparticles at $x=0$.
Due to the charge conservation there is, in fact, the creation of
particle-antiparticle pairs. Since the potential in $x>0$ is repulsive for
particles they are necessarily reflected. From the previous discussion
related to the classes B and C, one can realize that the threshold energy
for the pair production is given by $V=V_{c}$ for $c_{A}>1/2$ and that for $%
c_{A}\leq 1/2$ the pair production is not feasible. Evidently, the scalar
coupling increases the minimal energy necessary for the pair production. The
minimum value for the threshold ($V=2m$) occurs when there is a pure vector
coupling ($c_{A}=1$). The addition of a scalar contaminant contributes for
increasing the threshold, which surprisingly becomes infinity for a
half-and-half admixture of couplings. Then, the pair production is not
workable if the vector coupling does not exceed the scalar one, even if the $%
V$ is extremely strong. The propagation of antiparticles inside the
potential barrier can be interpreted as due to the fact that each
antiparticle is under the influence of an effective potential given by $%
\left( c_{S}-c_{A}\right) V$. In this way, each antiparticle has an
available energy (rest energy plus kinetic energy) given by $\left(
2c_{A}-1\right) V-E$, accordingly one concludes about the threshold energy.
One can also say that the particles are under the influence of an ascending
step of height $\left( c_{S}+c_{A}\right) V$, and that the antiparticles are
under the influence of an effective step of height $\left(
c_{S}-c_{A}\right) V$, an ascending step (repulsive) if $c_{A}<1/2$ and an
descending step (attractive) if $c_{A}>1/2$. For $E-m<V<V_{c}$ (class B),
one has that $J^{0}\gtrless 0$ for $V\lessgtr E/c_{A} $, thus the evanescent
wave with $q=+i\left\vert q\right\vert $ ($q=-i\left\vert q\right\vert $) is
related to particles (antiparticles). One can say that there is a charge
polarization due to the vector potential. The maximum charge density for
antiparticles occurs for $V=V_{c}$ and beyond this value they are pulled
apart. For the class B, the charge density beyond the potential barrier is
proportional to $\exp \left( -2\left\vert q\right\vert x\right) $ so that
the uncertainty in the position in the region $x>0$, estimated as being the
value of $x$ that makes the charge density equal to $J^{0}\left( 0\right) /e$%
, \ is given by $\Delta x=1/\left( 2|q|\right) $. This uncertainty presents
the minimum value
\begin{equation}
\left( \Delta x\right) _{\min }=\frac{1}{2\left( m+c_{S}V\right) }
\label{Dx}
\end{equation}%
when $V$ becomes $V=E/c_{A}$. From this last result one can see that $\left(
\Delta x\right) _{\min }=\lambda /2$ ($\lambda =1/m$ is the Compton
wavelength) in the case of a pure vector potential ($c_{S}=0$). However, one
can conclude that $\left( \Delta x\right) _{\min }<\lambda /2$ in the case
of a vector potential contaminated with some scalar coupling. Furthermore,
the penetration of the boson into the region $x>0$ shrinks without limit
with increasing $V$. At first glance it seems that the uncertainty principle
dies away provided such a principle implies that it is impossible to
localize a particle into a region of space less than half of its Compton
wavelength (see, e.g., Greiner 1990; Strange 1998). This apparent
contradiction can be remedied by recurring to the concepts of effective mass
and effective Compton wavelength. Indeed, Eq. (\ref{Dx}) suggests that we
can define the effective mass as $m_{\mathtt{eff}}=m+c_{S}V$ in such a way
that $\left( \Delta x\right) _{\min }=\lambda _{\mathtt{eff}}/2$ and $\left(
\Delta p\right) _{\max }=m_{\mathtt{eff}}$, where the effective Compton
wavelength is defined as $\lambda _{\mathtt{eff}}=1/m_{\mathtt{eff}}$. It
means that the localization of the boson does not require any minimum value
in order to ensure the single-particle interpretation of the DKP equation.

\section{Conclusions}

We have explored the influence of scalar and vector interactions in the DKP
formalism. We have shown that the spin-1 sector of the DKP theory looks
formally like the spin-0 sector and that the scalar coupling makes the DKP
formalism not equivalent to the Klein-Gordon or to the Proca formalisms.
With proper boundary conditions imposed on the DKP spinor for a square step
potential, we have analyzed in a very simple way the effects due to a scalar
coupling on the particle-antiparticle production and on the localization of
bosons. Another important conclusion of our work is that with an acceptable
definition of the reflection and transmission coefficients the charge is not
violated, even if Klein's paradox shows its face, and that the localization
of the boson does not require any minimum value in the context of the
single-particle interpretation of the DKP equation.

\bigskip

\bigskip

\bigskip

\bigskip

\bigskip

\noindent \textbf{Acknowledgments }This work was supported in part by means
of funds provided by CAPES and CNPq. 


\newpage

\begin{thebibliography}{99}
\bibitem{} Cardoso, T.R., de Castro, A.S.: Rev. Bras. Ens. Fis. \textbf{29},
203 (2007)

\bibitem{} Cardoso, T.R., Castro, L.B., de Castro, A.S.: Phys. Lett. A
\textbf{372}, 5964 (2008)

\bibitem{} Chetouani, L.,\textit{\ }Merad, M., Boudjedaa, T., Lecheheb, A.:
Int. J. Theor. Phys. \textbf{43}, 1147 (2004)

\bibitem{} Greiner, W.: Relativistic Quantum Mechanics: Wave Equations.
Springer, Berlin (1990)

\bibitem{} Kemmer, N.: Proc. Roy. Soc. of London A \textbf{173}, 91 (1939)

\bibitem{} Krajcik, R.A., Nieto, M.M.: Phys. Rev. D 10, 4049 (1974)

\bibitem{} Lunardi, J.T., Pimentel, B.M., Teixeira, R.G., Valverde, J.S.:
Phys. Lett. A \textbf{268}, 165 (2000)

\bibitem{} Merad, M.: Int. J. Theor. Phys. \textbf{46}, 2105 (2007)

\bibitem{} Nedjadi, Y., Barret, R.C.: J. Phys. G \textbf{19}, 87 (1993)

\bibitem{} Nedjadi, Y., Barret, R.C.: J. Math. Phys. \textbf{35}, 4517 (1994)

\bibitem{} Nowakowski, M.: Phys. Lett. A \textbf{244}, 329 (1998)

\bibitem{} Strange, P.: Relativistic Quantum Mechanics with Applications in
Condensed Matter and Atomic Physics. Cambridge University Press, Cambridge
(1998)
\end{thebibliography}
\end{document}